\documentstyle[12pt]{article} 
\begin{document} 
\begin{center} 
 
\vspace{7 cm} 
 
{\huge Lyapunov Exponents in Random 
 
\vspace{0.2 cm} 
 
Boolean Networks} 
 
\vspace{1 cm} 
 
{\Large Bartolo Luque$^{1,2,*}$ and Ricard V. Sol\'e$^{2,3}$}  

\vspace{0.2 in} 
 
(1) Centro de Astrobiolog\'{\i}a (CAB)  
 
Ciencias del Espacio, INTA 
 
Carretera de Ajalvir km. 4 

28850 Torrej\'on de Ardoz, Madrid, Spain

\vspace{0.5 cm}

(2) Complex Systems Research Group 
 
Departament of  Physics, FEN 
 
Campus Nord, M\`odul B4, 08034 Barcelona, Spain 
 
Universitat Polit\'ecnica de Catalunya 
 
\vspace{0.5 cm} 
 
(3) Santa Fe Institute 
 
Hyde Park Road 1399 
 
Santa Fe, New Mexico 87501, USA 
 
\vspace{0.2 in}

\end{center} 
 
\baselineskip=5.5 mm 
 
\vspace{0.25 cm}

\begin{abstract}

\vspace{0.25 cm} 
 
A new order parameter approximation to Random Boolean 
Networks (RBN) is introduced, based on the concept of Boolean 
derivative. A statistical argument involving an annealed approximation is  
used, allowing to measure the order parameter in terms of the 
statistical properties of a random matrix. Using the same formalism,  
a Lyapunov exponent is calculated, allowing to provide the onset of 
damage spreading through the network and how sensitive it is to minimal
perturbations. Finally, the Lyapunov exponents are obtained by means of
different approximations: through distance method and a discrete 
variant of the Wolf's method for continuous systems.
 
\end{abstract} 
 
\vspace{0.5 cm}

{\small $^*$ Corresponding author. E-mail: bartolo@complex.upc.es.} 

\begin{center} 
 
{\Large Submitted to Physica A}  
 
\vspace{0.2 cm} 
 
\end{center}

\newpage 
 
\section{Introduction} 
 
Random Boolean networks (RBN), also called Kauffman nets [1-2], were  
originally formulated as a model of genetic nets. The computational and
mathematical problems arising from continuous dynamical systems with a
very high number of coupled non-linear equations lead to the 
introduction of RBN as an alternative approach. In this way, relevant
statistical properties of RBN were derived [18].  This
general analysis allowed to test several hypothesis concerning the
large-scale organization of biological regulatory networks. 

Recent studies in the field try to obtain a natural bridge between discrete
and more biologically sensible, continuous networks [3]. In this vein for instance, a characteristic
quantitative measure associated to continuous dynamical systems is the
Lyapunov exponent $\lambda$ [4]. This parameter
measures the degree of instability of continuous dynamical systems,
and allows us to characterize the transition to chaos by measuring  
the pace at which initial conditions tend to diverge as the system evolves. 
Although Lyapunov exponents have been
also derived (or estimated) for discrete systems (as cellular automata [5-8])
 there is, as far as we know, no study about this quantity in
RBN and related systems. To have a way to estimate Lyapunov exponents for RBN's would thus establish a
natural link between discrete and continuous systems.

The paper is organized as follows. First the RBN formalism and its order-disorder phase transition 
are introduced. Secondly, using the 
concept of the Boolean derivative [9] we propose a new order 
parameter for the RBN phase transition: the percent of 1's 
in the Jacobian matrix that represents the Boolean derivative of the system. 
We then define a Lyapunov exponent for the RBN and compare our results 
with the distance method [11]. Finally, a second possible order parameter 
(the self-distance) is introduced. This will allow us, through a discrete analog of Wolf's method 
for continuous systems,
to reobtain an expression for the Lyapunov exponent consistent with that previously found.

\section{Random Boolean Networks}

A RBN is a discrete system involving $N$ units/automata with two possible states of a boolean variable
$\{0,1\}$. Each automaton is 
randomly connected with exactly $K$ neighbors. The state of each unit is updated
by means of a Boolean function, also  
randomly chosen from the set of all the Boolean functions with $K$ binary 
arguments.  Once the neighborhood and functions have been chosen they are  
fixed in time (i. e. a quenched set is used).  
The RBN exhibit a second order 
transition: for $K \leq 2$ a frozen (ordered) phase is observed, while for $K > 2$  
a disordered phase sets in. A RBN is by definition a discrete ($N$ cells) deterministic
system with a finite number of states ($\{0,1\}$), and therefore periodic patterns are expected
after a maximum of $2^N$ steps. Thus, if we follow strictly the standard 
definition of low-dimensional deterministic chaos, chaotic behavior is not possible in these systems. 
Taking this into account, we will define chaotic behavior here through 
damage spreading [12-14]: a phase will be chaotic if damage spreading takes place, i. e. if changes  
caused by transient flips of a single unit propagate and grow until they reach a size comparable to that 
of the system. Thus our disordered phase will be 
called chaotic phase, analogously to continuous systems.
 
\begin{figure} 
\vspace{8 cm} 
\includegraphics{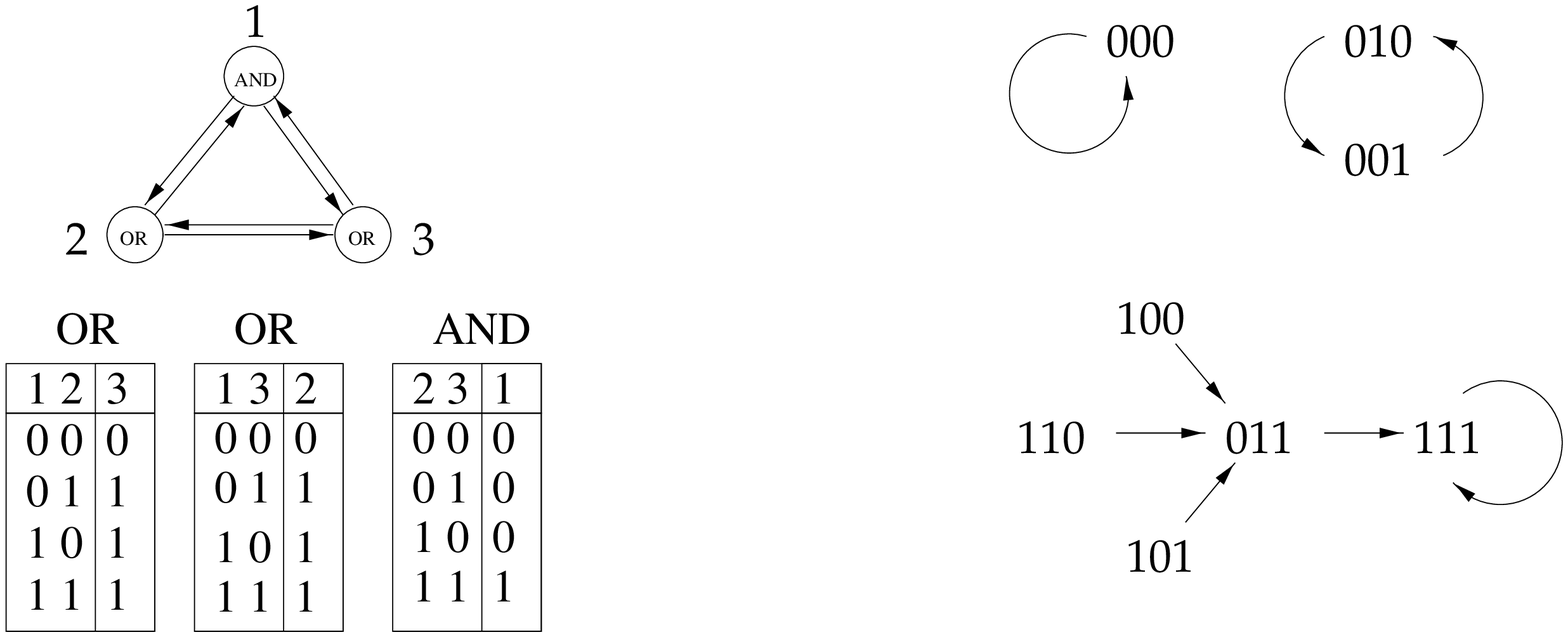} 
\caption{Example of RBN with $N=3$ and $K=2$. 
Left:  we show the interaction graph and the corresponding rule-tables 
(Boolean functions) of each automata. 
Right: we show explicitly the transitions between global system states 
as flow diagrams.} 
\end{figure} 

Let us illustrate the RBN structure and dynamics with a simple example [2] 
wich will be used below. 
Given a system with $N=3$ automata  
with values: $x_1, x_2$ y $x_3$ and connectivity $K=2$, the net is wired 
by choosing the input neighbors as indicated in figure 1.
 
The inputs have been represented in figure 1 as arrows that connect 
the automata and a simple directed
graph is obtained. The state of the system at any time is an ordered array of bits, 
 ${\bf X} = (x_1, x_2, x_3)$, and the interactions between the automatas 
are described by Boolean functions. 
In this example the Boolean functions have been
randomly sampled from the set of $2^{2^2}=16$ 
possible functions with $K=2$ arguments:  
 
\ 
 
(1) For the automaton 1: $f_1(x_2,x_3)$ the function AND. 
 
\ 
 
(2) For the automaton 2: $f_2(x_1,x_3)$ the function OR. 
 
\ 
 
(3) For the automaton 3: $f_3(x_1,x_2)$ the function AND. 
 
\ 
 
These functions are represented in figure 1 (left) by means of 
rule-tables (all possible inputs with their corresponding outputs). The system is 
updated synchronously.   
A possible temporal succession of states will describe a 
trajectory (orbit) of the system.   
In figure 1 (right) we represent all possible trajectories of the 
 system as a flow diagram.

 In early studies on RBN phase transitions the critical point was  
 estimated through numerical simulations [1,2]. The critical connectivity $K=2$  
gave the transition order-chaos. Later on this transition point was  analytically obtained by  
means of the so-called Derrida's annealed approximation [6,11], also 
known as the distance method. Derrida developed a non-correlated (annealed) RBN model by randomizing the inputs 
and Boolean functions at each time step showing that, in the thermodynamic 
limit, the transition point is the same in the quenched and the annealed  
systems. This approach can be extended to a continuous (average) $K$-valued
and biased RBN  [15,16]. In biased RBN's the Boolean functions are chosen with a bias $p$, that is: 
the mean percentage of 1's  in the output is $p$.
From two replicas with an initial normalized Hamming distance, $d(t=0)$, 
we can derive the equation for the evolution of $d(t)$:  
$$d(t+1) = 2p(1-p) \{ 1 -[1-d(t)]^{K} \} \eqno (1)$$ 

At the frozen phase, the fixed point $d^{*}=0$ is stable (i.e. the two 
initial configurations become identical as they evolve). In the chaotic phase however $d^{*}=0$ is 
unstable, and two initially close configurations diverge to a finite distance. 
The critical curve on the parameter space $(p,K)$ is: 
$$ K = {1 \over 2 p (1-p) } \eqno (2)$$ 
\begin{figure} 
\vspace{8 cm} 
\includegraphics{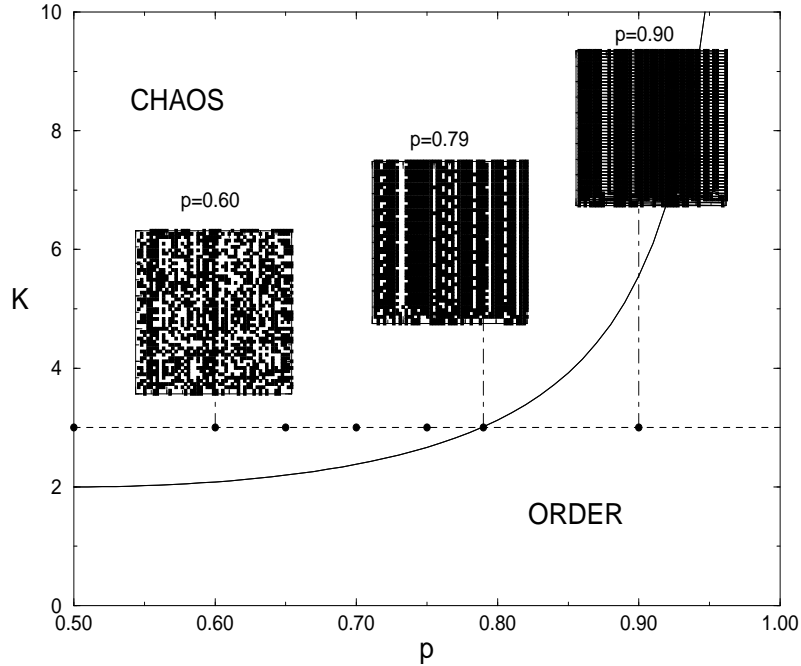}
\vspace{1 cm} 
\caption{Phase space for RBN. The critical line (continuous) is given by equation
2. It separates the ordered and chaotic phases. For $K=3$ (dashed line), 
three particular examples are shown for $N=50$ and random initial conditions.
Here $S_i=1,0$ are indicated as black and white squares, respectively.
In all space-time diagrams, time runs from bottom to top. The three
diagrams correspond to: $p=0.60$ (chaotic phase), $p=0.90$ (ordered phase) and
$p=0.79$, at the transition line.} 
\end{figure} 

In figure (2) the phase space is shown. On a constant connectivity line (here
$K=3$) three different runs of the system have been chosen for a RBN with
$K=3$ and $N=50$. For $p=1/2$ equation (2) reduces to the standard RBN problem [2]. 
Fig. 3 shows with continuous lines the evolution of $d(t)$ towards the theoretical fixed 
point $d^{*}$ (obtained from iteration of Eq. (1) with $K=3$), $p$ changing for each line. 
The values chosen for $p$ in this figure match the dots on the dashed $K=3$ line in Fig 2. 
The squares represent the average result of $100$ runs with two different replicas each. 
Here we used $N=10.000$ automata, and the initial distance is $d(0)=0.5$.  
Observe that $d^{*}$ acts as an order parameter. 
In fig. 4 the continuous line represents the stationary values obtained by iteration 
of equation (1) for changing $p$. Again, the squares represent the numerical values obtained by averaging 
over $100$ runs with two different replicas of RBN each, with size $N=10.000$, and $d(0)=0.50$.
\begin{figure} 
\vspace{8 cm} 
\includegraphics{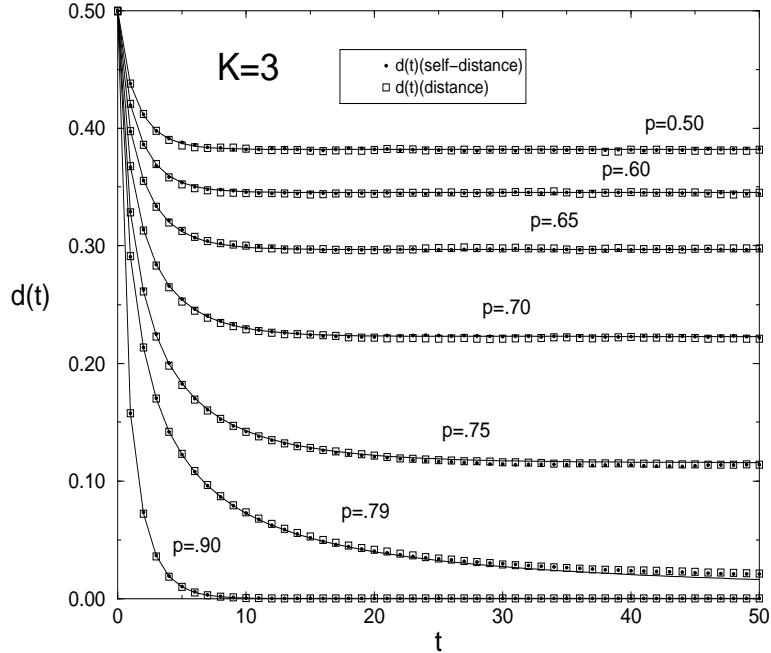} 
\vspace{1 cm} 
\caption{Continuous lines: dynamical evolution of the distance between 
two identical replicas of RBN with initial distance of $=.5$ for $K=3$ and different values of the bias $p$ through the iteration of equation (1).
Squares (points): numerical averages of the distance (auto-distance) for $100$ experiments with RBN of size $N=10.000$, $K=3$ and the bias showed.} 
\end{figure} 
 
In [17], Flyvbjerg defined a different order parameter:  
he defined the stable core at time $t$ as the set of units  
that have reached stable values at time $t$ (that is, remain unaltered in value for $t'\ge t$ and are  
independent of the initial conditions). Let us define $s(t)$ as the relative 
size of the  
stable core at time $t$, i.e. $s(t) N$ is its absolute value. Then the  
asymptotic stable core size, $s^*=\lim_{t\to\infty} s(t)$, is an 
order parameter for the order-chaos transition in RBN's.  
Flyvbjerg obtains an iterated equation for the stable core: 
$$ s(t+1) = \sum_{i=0}^{K} {K\choose i} s(t)^{K-i}  
(1-s(t))^i p_i  \eqno (3) $$ 
where $p_i$ is the probability that the Boolean function output be independent of a  
certain number $i$ of inputs. For the Boolean functions  
with bias $p$ it yields 
$p_i = p^{2^i} + (1-p)^{2^i}$. By analyzing the stability of Eq. (3),  
he found out an  
identical transition curve than the one given by Eq. (2). 
In Fig. 4 (Short-dashed line) we represent this order parameter as $1-s^{*}$.

\section{Boolean Derivatives in RBN}

We will now define a RBN in a more formal way. 
A RBN is a discrete dynamical system whose evolution is given by  
the iteration of a global mapping: 
$$ {\bf F}_{K,p}:\{0,1\}^N\mapsto \{0,1\}^N \eqno (4)$$  
where ${\bf F}_{K,p}=(f_1,f_2,...,f_N)$, and with each $f_i$ being a Boolean function of $K$ 
arguments and bias $p$ (mean percentage of ones in the outputs):   
$$ f_{K,p}:\{0,1\}^{K}\mapsto \{0,1\}_p \eqno (5)$$ 
A given configuration at time $t$, ${\bf x}(t)$, is updated sincrhonously, i.e.: 
$$ {\bf x}^{t+1} = {\bf F}_{K,p}({\bf x}^{t}) \eqno (6) $$ 
where each automata with $ x_i^t \in \{0,1\}$ is updated by mean of 
its corresponding Boolean function: 
$$ x_i^{t+1} = f_i(x_{i_1}^t,x_{i_2}^t, ... ,x_{i_K}^t) \eqno (7)$$ 
 
For a given ${\bf F}_{K,p}(t)$ we define its $N\times N$ Jacobian matrix, ${\bf F}_{K,p}^\prime(t)$, 
as that whose elements are given by the Boolean derivatives  
at time $t$ [9]: 
$$ {F}_{i,j}^\prime (t) = 
{\partial f_i(x_i^{t}) \over \partial {x_j}^t }= 
 \cases{ f_i(x_{i_1}^t,...,\bar {x_j}^t, ... ,x_{i_K}^t)\oplus  
f_i(x_{i_1}^t,...,{x_j}^t, ... ,x_{i_K}^t) &if ${x_j}$ inputs ${x_i}$ \cr 
0 &otherwise \cr} \eqno (8)$$ 
Here $\oplus$ is the exclusive OR (XOR) Boolean operation and  
$\bar x_j = x_j\oplus 1$ (i.e., the binary complement of $x_j$). 
From the point of view of damage spreading [7], 
 we can see that  
${F}_{i,j}^\prime(t) = 1$ if a flip in the input $x_j^t$ at 
time $t$ generates a change of
$x_{i}^{t+1}$ to $\bar x_{i}^{t+1}$ in step $t+1$. 
 In others words, the function 
spreads the damage. Otherwise, ${F}_{i,j}^\prime(t)= 0$, and no damage is  
spread. Note that ${\bf F}_{K,p}^\prime(t)$ depends on $t$  
because its 
value depends on the concrete configuration at time $t$.

Continuing with our example (see Fig 1), let us suppose that, at $t$, the state of the system is 
 ${\bf x}(t)=(1,0,1)$. If we compute the  $3 \times 3$ Jacobian matrix ${\bf F}_{2,0.4}^\prime(t)$ 
its components will be: 
$$F_{1,1}^\prime = 0$$ 
$$F_{1,2}^\prime = f_1(0,1) \oplus f_1(1,1) = 0 \oplus 1 = 1$$ 
$$F_{1,3}^\prime = f_1(0,1) \oplus f_1(0,0) = 0 \oplus 0 = 0$$ 
$$F_{2,1}^\prime = f_2(1,1) \oplus f_2(0,1) = 1 \oplus 1 = 0$$ 
$$F_{2,2}^\prime = 0$$ 
$$F_{2,3}^\prime = f_2(1,1) \oplus f_2(1,0) = 1 \oplus 1 = 0$$ 
$$F_{3,1}^\prime = f_3(1,0) \oplus f_3(0,0) = 1 \oplus 0 = 1$$ 
$$F_{3,2}^\prime = f_3(1,0) \oplus f_1(1,1) = 1 \oplus 1 = 0$$ 
$$F_{3,3}^\prime = 0$$ 
Thus the Jacobian matrix is: 
$${\bf F}^\prime(t)=\pmatrix {0&1&0\cr 0&0&0\cr 1&0&0\cr}$$ 
 
Analogously to the continuous dynamical systems counterpart  
we take a configuration state  
${\bf x}(t)$ and a slight perturbation of it ${\bf y}(t)$.  
A perturbed configuration is a new configuration at a non-zero (but otherwise small)  
Hamming distance from the original one. In fact, it is possible to define the perturbation ${\bf d}(t)$ 
such that:   
$${\bf y}(t)= {\bf x}(t)\oplus  {\bf d}(t) \eqno(9)$$ 
where the normalized Hamming distance between ${\bf x}(t)$ and ${\bf y}(t)$ 
is  
$$ \mid{\bf d}(t)\mid = {1\over N} \sum_{i=1}^N d_{i}^t \eqno (10)$$  
In our example, we have ${\bf x}(t)=(1,0,1)$ and we will take as the perturbed 
configuration ${\bf y}(t)=(0,0,1)$. Thus: 
$${\bf y}(t)= (0,0,1) = {\bf x}(t)\oplus  {\bf d}(t) = (1,0,1) 
 \oplus (1,0,0)=(1\oplus1,0\oplus0,1\oplus0)=(0,0,1)$$ 
Note that we have the minimum possible perturbation 
$$\mid{\bf d}(t)\mid ={1\over 3}(0+0+1)=1$$ 
and that we can  write the perturbation as 
$${\bf d}(t)= {\bf x}(t)\oplus  {\bf y}(t) \eqno (11)$$ 
Thus, in our example:  
$$(1,0,1) \oplus (0,0,1)=(1\oplus0,0\oplus0,1\oplus1)=(1,0,0)$$ 
Now we are ready to find the approximate evolution of the perturbation, 
as it is done in continuous systems. Using the equations (6), (9), (11), and 
using the Jacobian matrix to make a linear approximation, we have:  
$$ {\bf d}(t+1)= {\bf y}(t+1)\oplus {\bf x}(t+1) 
={\bf F} ({\bf y}(t))\oplus {\bf F} ({\bf x}(t))=$$ 
$$ ={\bf F} ({\bf x}(t) \oplus {\bf d}(t)) \oplus {\bf F} ({\bf x}(t)) 
\approx {\bf F}^\prime({\bf x}(t))\odot {\bf d}(t) \eqno (12)$$ 
where we define $\odot$ as: 
$$ {\bf F}^\prime ({\bf x}(t))\odot {\bf d}(t) =   
{\bf\Theta} ({\bf F}^\prime ({\bf x}(t))\cdot {\bf d}(t)) \eqno (13)$$  
and the vector ${\bf\Theta}({\bf x}) = ( \Theta(x_{1}),...,\Theta(x_{i}), ... ,\Theta(x_{N}))$ having 
standard Heaviside functions as components:  
$$ \Theta(x_{i}) = \cases {0 & if $x_{i}=0$ \cr  
                           1 & otherwise \cr} \eqno (14)$$ 
Thus, in our example we have:  
$$ {\bf d}(t+1)= \pmatrix {0&1&0\cr 0&0&0\cr 1&0&0\cr} \odot \pmatrix {1\cr 0\cr 0\cr} = 
{\bf\Theta}\Biggl[\pmatrix {0&1&0\cr 0&0&0\cr 1&0&0\cr} \cdot \pmatrix 
{1\cr 0\cr 0\cr} \Biggr] = \pmatrix {0\cr 0\cr 1\cr}$$  

\section{Jacobi matrix and a new order parameter} 
Our aim is now to present a new order parameter based on the Jacobian matrix of Boolean derivatives, 
as previously defined. The dynamical equation (12) can be  
iterated in $t$ to  
determine the evolution of the perturbation with two possible outcomes: as $t \to \infty$ either the 
initial perturbation at $t=0$ will tend to disappear, $\mid {\bf d}(\infty) \mid \to 0$, 
or it will reach a finite value.  
The behavior of the perturbation will be determined by the successive  
products of the Jacobian matrix. 
Thus, we define: 
 $${\bf M}(t)={\bf F}^\prime ({\bf x}(0)) 
\odot{\bf F}^\prime ({\bf x}(1))\odot \dots  
\odot{\bf F}^\prime ({\bf x}(t)) \eqno(15)$$ 
If the number of 1's in the Jacobian matrix is small the product in (15) will 
converge to a matrix ${\bf M}^{*}$ formed only by zeros, and 
 any initial perturbation will 
disappear. We will now attempt to construct an iterated equation for the evolution of the 
fraction of zeroes in the matrix $\bf M$ using a mean field approach. 
If our system has a connectivity $K$ and bias $p$, 
we substitute at each time step (mean field approximation) 
 the deterministic matrix   
${\bf F}^\prime (t)$ by a random matrix $\bf \Omega$ of the same form  
(at most $K$ 
1's at each row). The probability for a randomly generated Boolean function  
to have a Boolean derivative with value one ${\bf F}_{i,j}^\prime = 1$ 
is equal to the probability that  
a flip in its input $x_{i_j}$ generates a change in its output 
$x_{i}$ to $\bar x_{i}$. We have two possibilities: the output has a value 1 
and changes to 0, with probability $p(1-p)$ and the symmetric case with 
probability $(1-p)p$. Thus the mean number of 1's per row is $2p(1-p)K$.  
At each time step, $t+1$, we multiplie ${\bf M}(t)$ by a random matrix  
with a mean number  
$2p(1-p)K$ of 1's per row, i.e. ${\bf M}(t+1)=\bf\Omega{\bf M}(t)$.  

\begin{figure} 
\vspace{8 cm} 
\includegraphics{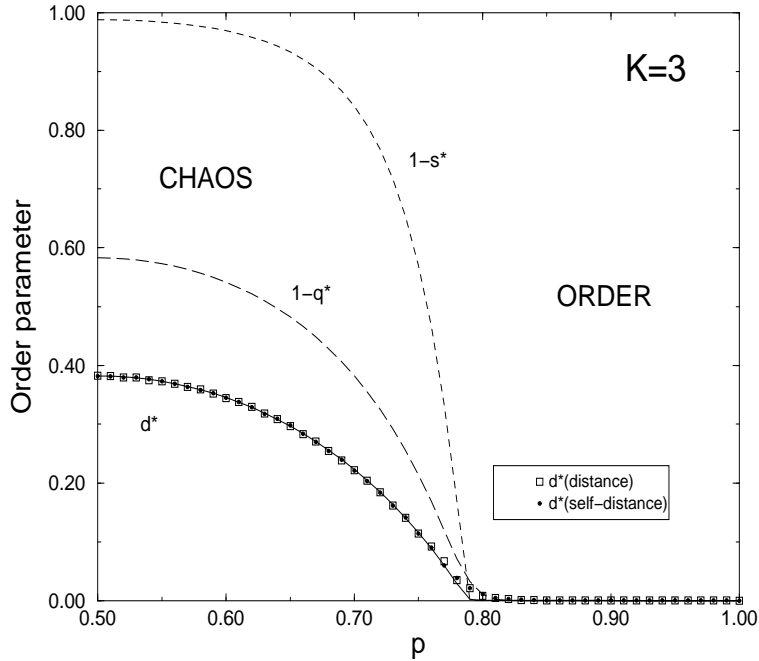} 
\vspace{1 cm} 
\caption{Continuous line: $d^*$ are assymptotic distances reached by 
iteration of equation (2) for different  $p$ with $K=3$. Short-dashed 
line: $s^*$ is the assymptotic unitary percent of elements of the stable
core reached by iteration of equation (3) by changing  $p$ with $K=3$. 
Long-dashed line: $1-q^*$ is the asymptotic unitary percent of ones in the 
matrix $M^*$ by iteration of equation (16).} 
\end{figure} 
This approach is clearly analogous 
to the distance method [11], where the Boolean functions and inputs of the  
system vary at random at each time step. So, if we assume that at a time  
step $t$ 
the matrix ${\bf M}(t)$ has a percentage $q_t$ of zeros, 
in the thermodynamic limit, the $q_{t+1}$ will be: 
$$q_{t+1}=\lim_{N\to\infty}\left[1-{2p(1-p)K \over N}(1-q_t)\right]^N 
=e^{-(1-q_t)2p(1-p)K} \eqno(16)$$   
Where $2p(1-p)K /N$ is the probability that $\Omega_{ij}=1$ and $1-q_t$ is the probability that $M_{jk}=1$.
By analyzing the stability of (16) around $q^{*}=1$ (i.e. all the elements of ${\bf M}^*$ are zero), 
we find out the critical transition curve, given again by 
$$ \left. {\partial q_{t+1} \over \partial q_t }  
\right\vert_{q^* = 1} = K 2p (1-p) <1 \eqno (17)$$ 
in agreement with previous results [7,11,17]. 
In Fig. 4 we indicate (long-dashed line) this order parameter as $1-q^{*}$ (the percentage of 1's in the infinite product of Jacobians) for $K=3$. 
\begin{figure} 
\vspace{8 cm} 
\includegraphics{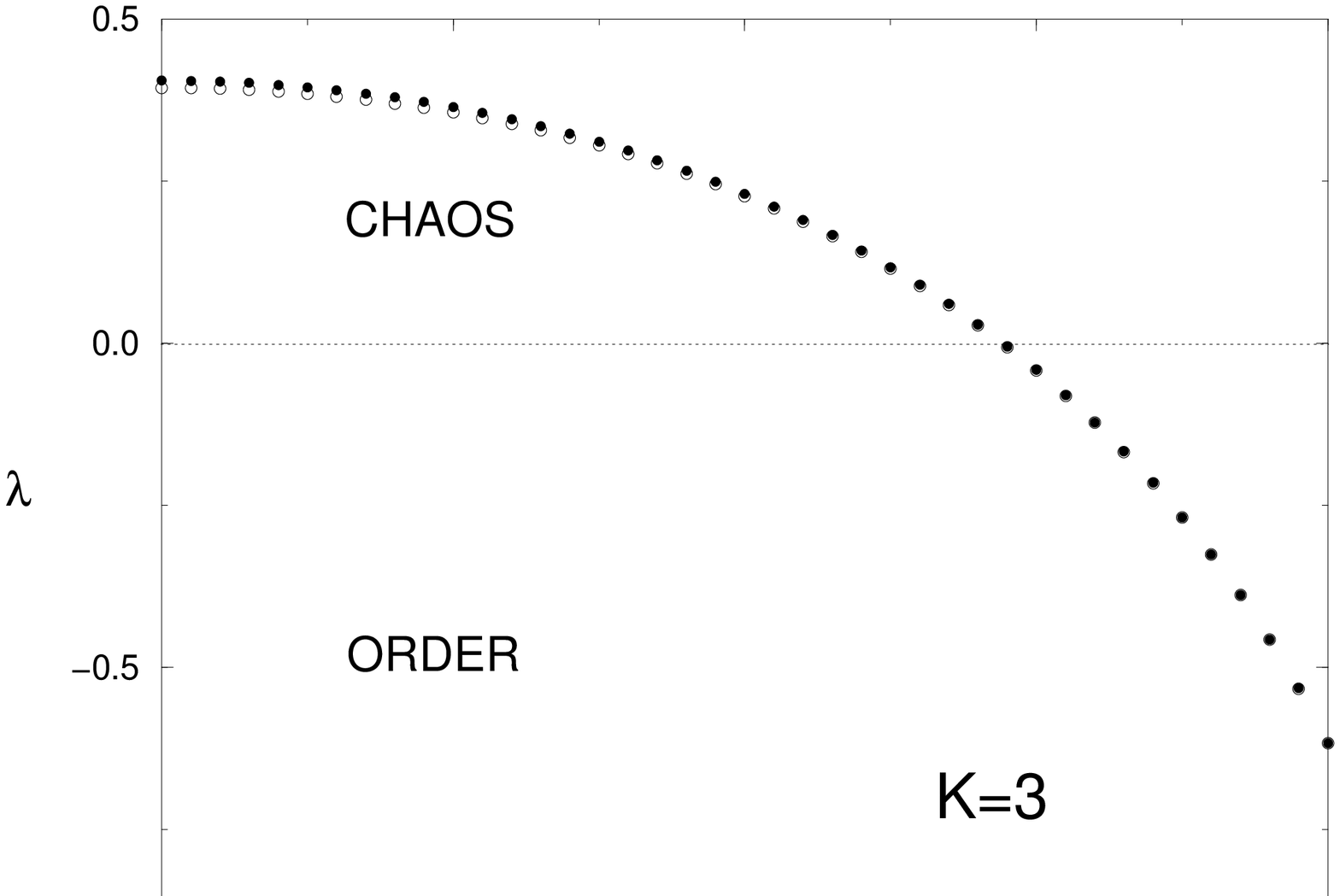} 
\vspace{1 cm} 
\caption{Open circles: values of Lyapunov exponent define from equation (27) 
with $K=3$ and different $p$ variating. Filled circles: numerical estimation
of the Lyapunov exponent by avering the expansion rate (18)
 calculate through the distance equation (1) with $T=10$ in equation (19).} 
\end{figure}
\section{Lyapunov exponents and RBN}

In the previous section, we have introduced the Boolean derivative  
${\bf F}^\prime ({\bf x}(t))$, in analogy with the standard continuous 
counterpart. This operator was then used in order to define the discrete  
map (12) which gives us the time evolution of the perturbation ${\bf d}(t)$.  
Using this definition, an expansion rate of perturbations for RBN can be  
easily defined. The damage expansion rate will be [6]: 
$$ \eta(t) = {\mid{\bf d}(t+1)\mid \over \mid{\bf d}(t)\mid} \eqno (18)$$ 
This allows us to define a Lyapunov exponent: 
$$ \lambda (T) = {1\over T} \sum_{t=1}^T \log \eta (t) \eqno (19)$$  
Under the previous approach we can determine the mean damage expansion 
rate $\bar{\eta}$ which will be given by (here $\langle ... \rangle$ are  
time averages): 
$$\bar{\eta}=\langle\eta(t)\rangle \eqno (20)$$ 
$$= \Bigg\langle{\mid{\bf d}(t+1) 
\mid \over \mid{\bf d}(t)\mid}\Bigg \rangle \eqno (21)$$ 
$$= \Bigg\langle{\mid{\bf F}^\prime({\bf x}(t))\odot {\bf d}(t) 
\mid \over \mid{\bf d}(t)\mid}\Bigg \rangle \eqno (22)$$ 
such quantity can be easily computed by analyzing the statistical behavior 
of $\mid{\bf F}^\prime({\bf x}(t))\odot {\bf d}(t)\mid$. This can be done  
by assuming that, on mean field grounds, ${\bf F}^\prime ({\bf x}(t))$ can    
be replaced by a random matrix $\bf\Omega$. The previous average (19) can  
then be  
estimated by considering the percent of 1's in ${\bf d}(t)$ (i.e. 
$\mid{\bf d}(t)\mid / N$) and the same quantity for $t+1$ (i.e. 
$\mid{\bf d}(t+1)\mid / N$). We have: 
$${\mid{\bf d}(t+1)\mid \over  N} = 1 - \left[1-{2p(1-p)K \over N} 
{\mid{\bf d}(t)\mid \over  N}\right]^N \eqno (23)$$ 
Now, in the thermodynamic limit ($N\to\infty$) we get: 
$$ \bar{\eta}=   1 -  
\exp \Biggl [{-2p(1-p)K {\mid{\bf d}(t)\mid \over  N} \over 
{\mid{\bf d}(t)\mid \over  N}}  
\Biggr ] \approx 2p(1-p)K \eqno(24)$$ 
This result could be derived in another way. By defining the normalized 
Hamming distance of a Boolean matrix $\bf \Omega$ as: 
$$ \mid{\bf \Omega}\mid = {1\over N^2} \sum_{i,j}^N \Omega_{ij} \eqno (25)$$  
where $\Omega_{ij}\in \{0,1\}$ 
We have: 
$$\bar{\eta}(t)= {\mid{\bf \Omega}\odot {\bf d}(t) 
\mid \over \mid{\bf d}(t)\mid}={\mid{\bf \Omega}\mid \mid{\bf d}(t) 
\mid \over \mid{\bf d}(t)\mid} = \mid{\bf \Omega}\mid= 2p(1-p)K\eqno (26)$$ 
From (19) and (24), the Lyapunov exponent will be: 
$$ \lambda = \log \Bigl [ 2p(1-p)K \Bigr ] \eqno(27)$$ 
which determine the two classical regimes: $\lambda <0$ (order) and  
$\lambda > 0$ (chaos) with the marginal case $\lambda = 0$. 
In agreement with the boundary phase transition (2).

\section{Distance and Wolf's method }

This result has been consistent with the equation of distance evolution (1). 
If we interpret one of the replicas in the distance method as a  
perturbation of the another replica, the expansion in the time $t$ of 
 the perturbation will be:   
$$ \eta(t) = {\mid d_{12}(t+1)\mid \over 
 \mid d_{12}(t)\mid}= {2p(1-p)\{1-[1-d(t)]^K\}\over d(t)} \eqno (28)$$ 
and approximating for small $d(t)$: 
$${(1-d(t))^K} \approx {1-Kd(t)} \eqno (29)$$ 
we have: 
$${\eta(t)} \approx {2p(1-p)K} \eqno (30)$$  
i.e an expansion rate identically to (24). 

This approximation is prove sufficiently good as we show in fig. 5. The values
of Lyapunov exponents through equation (27) 
with $K=3$ and $p$ variating and there equivalents values calculated through
 the distance equation (1) coincides.

There are different methods to compute Lyapunov exponents in continuous systems [4]. 
 In order to show consistence we will demonstrate that it is possible 
 computate Lyapunov exponents from self-distance in consonance 
 with the previous result. 
 
The Wolf's method is used to numerically estimate Lyapunov  
exponents from time series. In short, the method is as follow: get two points 
 of the time series, let us say ${\bf X}(t_1)$ and ${\bf X}(t_2)$ and 
 compute their relative distance: $\mid {\bf X}(t_2) - {\bf X}(t_1) \mid$. 
 Assume that $\mid {\bf X}(t_2) - {\bf X}(t_1) \mid < \epsilon$,  
being $\epsilon > 0$ very small. Next, compute the distance after $T$ steps,
 i.e: $\mid {\bf X}(t_2+T) - {\bf X}(t_1+T) \mid$. 
  This time $T$ is a fraction of the characteristic period or  
is defined in terms of the autocorrelation function. 
 Repeating for $n$ pairs of points and averaging, we obtain an estimation 
 of the Lyapunov exponent: 
$$\lambda = {1 \over nT} \sum_{t_{2} \neq t_{1}}^{n} 
{ \log{ { \mid {\bf X}(t_2+T) - {\bf X}(t_1+T) \mid \over \mid {\bf X}(t_2) - 
 {\bf X}(t_1) \mid } } } \eqno (31)$$ 
For RBN, we can write an equation for the normalized Hamming distance between 
successive time steps in our system, i.e. the self-distance: $d_{t,t-1}$. It easy to
see that the self-distance is a new order parameter. This is a consequence of the combination 
of the distance method and the stable core.

The iterated equation for the self-distance is: 
$$ d_{t+1,t} = 2p(1-p)[1-(1-d_{t,t-1})^K]  \eqno (32)$$
Wich formally is equivalent to equation (1) but different conceptually.
The self-distance, as the stable core, does not require annealed replicas
(as the distance method), and is computationally more easily to determine.
In fig. 3-4 the numerical values of self-distance (points) are calculated in similar way
that the distance (squares) with a very good agreement.
 
If we approximate linearly close to the fixed point $d^*=0$, the function becomes: 
$$ d_{t+1,t} =  2p(1-p)Kd_{t,t-1}  \eqno (33)$$ 
The iterated equation now is resoluble: 
$$ d_{t+T,t} =  [ 2p(1-p)K ]^T d_{t,t-1}  \eqno (34)$$ 
Thus, we have: 
$${d_{t_2+T}-d_{t_1+T} \over d_{t_2}-d_{t_1}}=[K2p(1-p) ]^T
\eqno (35)$$ 
i. e. aconstant value that, after introduced in the sum of (31), gives the  Lyapunov 
exponents (27).

\section{Summary}
 
In this paper we have analyzed a new order parameter for RBN in  
terms of a $\bf\Omega$-random matrix approach. Our order parameter deals
 with the percent of non-zero  
elements that is obtained from the limit $\lim_{t\to\infty}{\bf\Omega}^t$.   
It is shown that the order parameter describes a (second order) phase  
transition at a critical point consistent with other previous analyses. 
 
An inmediate extension of the Boolean derivative approach is the construction 
of a measure for the damage expansion rate, $\eta(t)$ that give a  
quantitative characterization of how small perturbations propagate  
through the network. It has been shown that $\eta(t)$ provides a consistent 
measure of such sensitivity to spin flips. The time average over the Boolean 
products of the Jacobi matrix on the distance vectors can be successfully  
translated to a simple annealed method where only the statistical properties  
of the random matrix $\bf \Omega$ matter. 

We also have calculated the Lyapunov exponents through the distance method.
And we propose a new order parameter: the self-distance. This quantity opens the 
possibility of defining the Lyapunov exponent in a discrete system in analogy 
with the Wolf's method for continuous systems and the result is in 
agreement with the previously obtained.

\vspace{1 cm} 
 
{\section {Acknowledgments} 
 
\vspace{0.5 cm} 
 
The authors would like to thank Antonio Ferrera and Stuart Kauffman for help at different stages 
of this study. This work has been  
partially supported by a grant DGYCIT PB-97-0693 and by the the Santa Fe Institute (RVS) and by the Centro de Astrobiolog\'{\i}a (BLS).

\newpage 
 
\section{References} 

\noindent 
[1] S. A. Kauffman, J. Theor. Biol., {\bf 22}:437 (1969) 
 
\vspace{0.2 cm} 

\noindent 
[2] S. A. Kauffman, The Origins of Order, Oxford U. Press (Oxford, 1993)
  
\vspace{0.2 cm}

\noindent
[3] T. Mestl, R.J. Bagley and L. Glass, Phys. Rev. Lett. 79 (1997) 653

\vspace{0.2 cm}

\noindent
[4] J. Guckenheimer and P. Holmes, Nonlinear oscillations, dynamical systems and
bifurcations of vectors fields, Springer (New York, 1982)

\vspace{0.2 cm}

\noindent
[5] S. Wolfram, Cellular automata and complexity. Adison Wesley (1994)

\vspace{0.2 cm} 

\noindent
[6] F. Bagnoli, R. Rechtman and S. Ruffo, Phys. Lett. A, {\bf 172}:34 (1992)

\vspace{0.2 cm}

\noindent
[7] J. Ur\'ias, R. Rechtman, A. Enciso. Chaos, 7, 688 (1997) 
 
\vspace{0.2cm} 

\noindent 
[8] Shereshevsky, J. nonlinear Sci. 2, 1 (1992) 
 
\vspace{0.2cm} 

\noindent 
[9] G. Y. Vichniac: {\it Physica D}, {\bf 45}:63 (1990) 
 
\vspace{0.2cm} 

\noindent 
[10] F. Bagnoli, Int. J. Mod. Phys. C 3, 307 (1992) 
 
\vspace{0.2cm}

\noindent
[11] B. Derrida and Y. Pomeau, Europhys. Lett., {\bf 1}:45 (1986) 
 
\vspace{0.2 cm} 

\noindent 
[12] D. Stauffer, in: Chaos and Complexity, eds. R. Livi et at. (World Scientific, Singapore, 1987) 
 
\vspace{0.2cm} 

\noindent 
[13] F. Bagnoli, J. Stat. Phys., {\bf 85}:151 (1996) 
 
\vspace{0.2cm} 

\noindent 
[14] B. Luque and R. V. Sol\'e, Phys. Rev. E55 (1997) 257  
 
\vspace{0.2 cm} 

\noindent 
[15] R. V. Sol\'e and B. Luque, Phys. Lett. A,{\bf 196}:331 (1995) 
 
\vspace{0.2cm}

\noindent 
[16] G. Weisbuch and D. Stauffer, J. Physique 48 (1987) 11-18 
 
\vspace{0.2cm} 
  
\noindent 
[17] H. Flyvbjerg J. Phys. A: Math. Gen., {\bf 21}:L955 (1988) 
 
\vspace{0.2 cm} 

\noindent 
[18] U. Bastolla and G. Parisi, Physica D 98 (1996) 1-25 739
\vspace{0.2cm}

\end{document}